\documentclass[aps,prd,10pt,twocolumn,nofootinbib]{revtex4-1}
\usepackage{graphicx, epsfig}
\usepackage{amsmath,amssymb,amsfonts,dsfont,mathrsfs,amsthm,mathtools}
\usepackage{bm}
\usepackage[normalem]{ulem}
\usepackage{color}
\usepackage[dvipsnames]{xcolor}
\usepackage[unicode]{hyperref}
\hypersetup{colorlinks=true, citecolor=MidnightBlue,
            linkcolor=MidnightBlue, urlcolor=MidnightBlue, linktocpage=true}
\usepackage{array}
\usepackage{booktabs}
\usepackage{blindtext}
\usepackage{amsfonts}
\usepackage{physics}
\usepackage{lipsum}

\newcommand{\Lag}{\mathscr{L}}

\def\Q{\mathcal{Q}}

\begin{document}

\title{Thermodynamics of Lorentzian Taub--NUT spacetimes in\\ Einstein--Gauss--Bonnet AdS gravity}

\author{Borja \surname{Diez}
}
\email{borja.diez@cinvestav.mx}
\affiliation{Departamento de F\'{\i}sica, Cinvestav, Av.~IPN 2508, 07360, CDMX, M\'exico}

\begin{abstract}
        We study the thermodynamics of Lorentzian Taub--NUT spacetimes in Einstein--Gauss--Bonnet AdS gravity in arbitrary even dimensions, constructed as $U(1)$ fibrations over Einstein--K\"ahler base manifolds. We compute the temperature from the surface gravity and the energy as the conserved charge associated with the stationary Killing vector using the off-shell Abbott--Deser--Tekin formalism. Within the same framework, we obtain the entropy as a horizon Noether charge. The result agrees with the Iyer--Wald formula and includes a Gauss--Bonnet correction to the Bekenstein--Hawking area law. Retaining the Misner strings whenever present, we formulate a first law of full cohomogeneity in which the NUT parameter varies independently of the horizon radius. This requires an additional NUT-related thermodynamic variable and its conjugate potential, which we determine explicitly. We also briefly discuss the static limit.

\end{abstract}

\maketitle

\section{Introduction}\label{sec:Introduction}

Black hole thermodynamics offers fundamental insights into the quantum nature of gravity~\cite{Bekenstein:1972tm,Hawking:1975vcx,Bekenstein:1973ur,Hawking:1976ra,Bardeen:1973gs}. In this framework, black holes behave as thermodynamic systems whose properties are expected to admit a microscopic description in terms of quantum states~\cite{Almheiri:2020cfm}. Identifying the degrees of freedom responsible for this behavior remains a central challenge in quantum gravity. Already at the semiclassical level, however, black holes obey thermodynamic laws that relate their geometric properties to temperature, entropy, and energy~\cite{Gibbons:1976pt,Hawking:1982dh,York:1986it,Hawking:1995fd,Gibbons:1976ue,Rathi:2021aaw}. Gravitational thermodynamics also extends beyond black holes to configurations with more intricate global and topological structures, including geometries without a conventional event horizon. Although the general principles of gravitational thermodynamics are well established, formulating thermodynamic laws for a specific spacetime can be challenging and requires a detailed analysis of the solution. To avoid degeneracies associated with restricted variations, the first law should have full cohomogeneity: its independent variations must span the space of independent parameters characterizing the family of solutions. This may require introducing additional thermodynamic variables and their conjugates. Examples include scalar charges~\cite{Gibbons:1996af}, thermodynamic volumes~\cite{Kastor:2009wy,Dolan:2010ha,Cvetic:2010jb}, thermodynamic lengths~\cite{Appels:2017xoe,Anabalon:2018qfv}, Misner charges~\cite{Bordo:2019tyh}, and magnetic moments~\cite{Gibbons:2013dna,Kubiznak:2026uro}.

Taub--Newman--Unti--Tamburino geometries, hereafter Taub--NUT spacetimes, were introduced as exact solutions of Einstein's equations that generalize the Schwarzschild geometry through an additional parameter known as the NUT charge~\cite{Taub:1950ez,Newman:1963yy}. This parameter can be interpreted as a gravitational analogue of magnetic charge~\cite{Plebanski:1975xfb,Liu:2022uox} and gives rise to a nontrivial fibration of the temporal coordinate over the base manifold. The resulting spacetime is stationary and, in the standard four-dimensional construction, axisymmetric. The NUT parameter also modifies the asymptotic geometry, which is no longer globally Minkowski or (anti-)de Sitter ((A)dS) in the usual sense. Although these geometries can be free of local curvature singularities, they may contain a Misner string~\cite{Misner:1963fr}, a string-like defect associated with the nontrivial time fibration. Conventionally, this string is removed by imposing an appropriate periodic identification of the temporal coordinate. Such an identification is not essential for geodesic completeness, however, which can be maintained in the presence of the Misner string~\cite{Clement:2015cxa,Clement:2015aka}. This reflects the global nature of the string, rather than a local curvature singularity obstructing geodesic motion. Although the NUT charge was originally introduced on theoretical grounds, its possible presence in astrophysical compact objects has also been investigated~\cite{Chakraborty:2017nfu,Chakraborty:2019rna,Ghasemi-Nodehi:2021ipd}, as has the potential role of NUT-charged primordial black holes in dark matter scenarios~\cite{Chakraborty:2022ltc,Chakraborty:2023wpz}. It has also been shown to affect the holographic DC conductivity~\cite{Khan:2025fne} and Hall transport~\cite{Khan:2026kcs}.

The global subtleties associated with the Misner string have historically motivated the study of Taub--NUT geometries in Euclidean signature~\cite{Misner:1963fr,Hawking:1998ct,Mann:1999pc,Mann:1999bt,Mann:2004mi,Astefanesei:2004kn,Liko:2011cq}. In this setting, they can be interpreted as gravitational instantons and contribute as nontrivial saddle points to the gravitational path integral~\cite{Gibbons:1979xm}. They therefore provide a setting for studying semiclassical quantum gravity, gravitational partition functions, and transitions between geometries with different topologies. The Euclidean version of this geometries can be characterized by the fixed-point set of the Killing vector associated with the Euclidean time $\xi=\partial_\tau$: a zero-dimensional fixed-point set is called a \textit{nut}, whereas a fixed-point set of codimension two is called a \textit{bolt}~\cite{Gibbons:1979xm}.

On the other hand, higher-curvature extensions of general relativity are motivated by several considerations. When gravity is treated as a Wilsonian effective field theory, higher-curvature terms arise naturally as corrections to the Einstein--Hilbert action. Such interactions also appear in the low-energy limit of string theory~\cite{Zwiebach:1985uq} and modify the properties of higher-dimensional black holes relative to their Einstein-gravity counterparts~\cite{Boulware:1985wk,Wheeler:1985nh,Wheeler:1985qd}. In quantum field theory on curved spacetime, renormalizing the matter stress-energy tensor requires higher-curvature counterterms in the gravitational action~\cite{Birrell:1982ix}. Curvature invariants also play a central role in quantum anomalies~\cite{Graham:1999pm,Alvarez-Gaume:1983ihn,Capper:1975ig,Deser:1993yx,Buchel:2003tz}. In four dimensions, adding suitable curvature-squared terms yields a theory that is perturbatively renormalizable around Minkowski spacetime~\cite{Stelle:1976gc}, although it also introduces ghost-like degrees of freedom~\cite{Stelle:1977ry}. Within the anti-de Sitter/conformal field theory (AdS/CFT) correspondence~\cite{Maldacena:1997re,Witten:1998qj,Gubser:1998bc}, higher-curvature interactions allow one to explore holography beyond Einstein--AdS gravity by modifying the holographic dictionary and encoding additional properties of the dual CFT~\cite{Sinha:2010ai,Alishahiha:2010bw,Kwon:2011jz,Cunliff:2013en,Ghodsi:2019xrx,Ghodsi:2020qqb,Anastasiou:2025dex,Anastasiou:2021swo}.

Among these extensions, Lovelock gravity~\cite{Lovelock:1971yv,Lanczos:1938sf} is distinguished by its second-order field equations. It is the most general purely metric theory constructed from the metric and the Riemann tensor with this property in arbitrary dimensions. Despite containing higher powers of the curvature, it therefore avoids the Ostrogradsky instabilities that typically arise in higher-curvature theories. Its action principle is a linear combination of dimensionally continued Euler densities, that is,
\begin{equation}\label{I-Lovelock}
\begin{split}
    I_{\rm Lovelock}[g]
    &=
    \sum_{p=0}^{\left[d/2\right]}
    \int_{\mathcal{M}}\dd^{d+1}x\,\sqrt{|g|}\alpha_p\Lag^{(p)}\,,\\
    \Lag^{(p)}&=\frac{1}{2^p}
    \delta^{\mu_1\ldots\mu_{2p}}_{\nu_1\ldots\nu_{2p}}
    R^{\nu_1\nu_2}_{\mu_1\mu_2}
    \cdots
    R^{\nu_{2p-1}\nu_{2p}}_{\mu_{2p-1}\mu_{2p}}\,,
\end{split}
\end{equation}
where $g$ is the determinant of the metric, $[\ldots]$ denotes the integer part, $\alpha_p$ are the Lovelock couplings, and we have defined the generalized Kronecker delta as
\begin{equation}
    \delta^{\mu_1\ldots\mu_p}_{\nu_1\ldots\nu_p}
    =
    p!\,\delta^{[\mu_1}_{[\nu_1}\cdots
    \delta^{\mu_p]}_{\nu_p]}\,.
\end{equation}
The first three terms correspond to the cosmological term, the Einstein--Hilbert term, and the Gauss--Bonnet density, respectively. Einstein--Gauss--Bonnet (EGB) gravity is thus the simplest higher-curvature extension within the Lovelock family, obtained by truncating the series at quadratic order in the curvature. Black hole thermodynamics in Lovelock gravity has attracted sustained interest as a means of understanding the effects of these interactions~\cite{Cai:2003kt,Khodam-Mohammadi:2008hww}.

The thermodynamics of Taub--NUT geometries has been studied in both Euclidean signature~\cite{Ciambelli:2020qny,Mann:1999bt,Mann:2004mi,Astefanesei:2004kn,Liko:2011cq,Khodam-Mohammadi:2008hww,Clarkson:2002uj,Astefanesei:2004ji} and the Lorentzian sector~\cite{Durka:2019ajz,Frodden:2021ces,Bordo:2019slw,BallonBordo:2019vrn,Hennigar:2019ive,Tharwat:2026ipi}. These approaches assign different thermodynamic roles to the NUT charge and consequently lead to different formulations of the first law. In the Euclidean approach, thermodynamic quantities are usually extracted from the renormalized on-shell action. This requires boundary terms that ensure a well-posed variational principle, together with counterterms that cancel divergences. In the standard intrinsic counterterm construction, the number and complexity of the required curvature terms increase with the spacetime dimension, making calculations in arbitrary dimensions progressively more involved~\cite{Clarkson:2002uj,Astefanesei:2004ji}.

An alternative method for computing the energy is provided by the off-shell Abbott--Deser--Tekin (ADT) formalism introduced in Ref.~\cite{Kim:2013zha}. This prescription reproduces conserved charges obtained through independent methods and also applies to geometries with nonstandard asymptotics, including spacetimes with anisotropic scaling~\cite{Ayon-Beato:2015jga,Gim:2014nba}. Its construction requires the bulk Lagrangian and a path in the space of solutions connecting a reference background to the configuration of interest, without explicitly introducing holographic counterterms or additional boundary terms. These features make it well suited to the study of thermodynamics properties of Taub--NUT spacetimes in higher dimensions.


In this work, we extend the Lorentzian full-cohomogeneity thermodynamic framework introduced in Ref.~\cite{Hennigar:2019ive} for Einstein gravity to Taub--NUT configurations in Einstein--Gauss--Bonnet AdS (EGBAdS) gravity. The underlying Taub--NUT solutions in higher-curvature gravity have been studied previously; our purpose here is instead to determine how the Gauss--Bonnet corrections modifies their Lorentzian thermodynamic description when the Misner strings are retained and the NUT parameter is allowed to vary independently of the horizon radius. We compute the energy using the off-shell ADT formalism and derive the entropy both as a horizon Noether charge and from the Iyer--Wald prescription. We then determine the additional NUT-related thermodynamic pair required for a first law of full cohomogeneity. Finally, we examine the static limit and compare the resulting thermodynamic quantities with those obtained through holographic renormalization for the six-dimensional topological Boulware--Deser black hole~\cite{Anastasiou:2025usa}.


The remainder of this manuscript is organized as follows. Sec.~\ref{sec:EGB gravity} introduces EGB gravity with a negative cosmological constant and specifies our conventions. In Sec.~\ref{sec:TN-EGB}, we present the Taub--NUT solution with Gauss--Bonnet corrections and discuss its geometric properties. Sec.~\ref{sec:Thermo} develops its Lorentzian thermodynamics, including the energy, entropy, temperature, and NUT-related thermodynamic pair, and establishes the first law. We also briefly examine the static limit of these configurations, obtained by taking the NUT parameter to zero, $n\to0$. Finally, Sec.~\ref{sec:Conclusions} summarizes our results and outlines directions for future research.

\section{Einstein--Gauss--Bonnet AdS gravity}\label{sec:EGB gravity}
EGB gravity is a particular Lovelock theory obtained by truncating the series in Eq.~\eqref{I-Lovelock} at quadratic order in the curvature. Its action reads
\begin{equation}\label{I-EGB}
    I_{\rm EGB}[g]
    =
    \kappa\int_{\mathcal{M}}\dd^{d+1}x\,\sqrt{|g|}
    \left(
        R+\frac{d(d-1)}{\ell^2}+\alpha\mathcal{G}
    \right)\,,
\end{equation}
where $\kappa=(16\pi G)^{-1}$, with $G$ denoting Newton's constant, and $\ell$ is the bare AdS radius, related to the cosmological constant by
\begin{equation}
    \ell^2=-\frac{d(d-1)}{2\Lambda}\,.
\end{equation}
The parameter $\alpha$ is the Gauss--Bonnet coupling, and
\begin{equation}
    \mathcal{G}=
    R_{\mu\nu\alpha\beta}R^{\mu\nu\alpha\beta}
    -4R_{\mu\nu}R^{\mu\nu}
    +R^2\,,
\end{equation}
is the Gauss--Bonnet density. In four spacetime dimensions, corresponding to $d=3$ in our conventions, this term is topological and does not contribute to the bulk equations of motion. Nevertheless, it plays an important role in the renormalization of the Euclidean action and the definition of conserved charges for asymptotically locally AdS spacetimes~\cite{Aros:1999id,Aros:1999kt,Mora:2004kb,Kofinas:2006hr,Olea:2006vd,Arenas-Henriquez:2019rph}. In particular, when evaluated on four-dimensional configurations with an (anti-)self-dual Weyl tensor, the topologically renormalized action becomes proportional to the Pontryagin index~\cite{Corral:2021xsu,Ciambelli:2020qny,Corral:2025npd}. This parallels the behavior of Yang--Mills instantons~\cite{Belavin:1975fg}, whose on-shell action is likewise governed by a topological invariant.

Varying the action in Eq.~\eqref{I-EGB} with respect to the metric yields
\begin{equation}\label{eom}
    R_{\mu\nu}
    -\frac{1}{2}g_{\mu\nu}R
    -\frac{d(d-1)}{2\ell^2}g_{\mu\nu}
    +\alpha H_{\mu\nu}
    =0\,,
\end{equation}
where
\begin{equation}
\begin{split}
    H^\mu{}_\nu
    &=
    2R^{\mu\rho}{}_{\sigma\lambda}
      R^{\sigma\lambda}{}_{\nu\rho}
    -4R^\sigma{}_\rho
      R^{\mu\rho}{}_{\nu\sigma}
    +2R R^\mu{}_\nu\\
    &~~
    -4R^\mu{}_\lambda R^\lambda{}_\nu
    -\frac{1}{2}\delta^\mu_\nu\mathcal{G}\,,
\end{split}
\end{equation}
is the contribution arising from the variation of the Gauss--Bonnet term with respect to the metric. Diffeomorphism invariance of the theory ensures that this tensor is identically covariantly conserved,
\begin{equation}
    \nabla_\mu H^\mu{}_\nu=0\,.
\end{equation}
Although the action~\eqref{I-EGB} is quadratic in the curvature, the field equations contain at most second derivatives of the metric for generic backgrounds, a defining property of Lovelock theory~\cite{Lovelock:1971yv}.

In more than four spacetime dimensions, the Gauss--Bonnet term contributes dynamically, making EGB gravity one of the simplest settings for analytically exploring higher-curvature effects while retaining second-order field equations. The theory admits exact static black hole solutions, first obtained by Boulware and Deser~\cite{Boulware:1985wk} and subsequently extended to different horizon topologies and asymptotic geometries~\cite{Wheeler:1985nh,Wheeler:1985qd,Cai:2001dz}. These solutions provide a useful setting for studying how higher-curvature interactions affect the causal structure, thermodynamics, stability, and phase structure of black holes.

EGB gravity also has a richer vacuum structure than Einstein gravity. Its maximally symmetric vacua are determined by a quadratic polynomial in the effective cosmological constant, generically yielding two branches with distinct effective curvature radii. One branch approaches Einstein gravity continuously as $\alpha\to 0$, whereas the other has no smooth Einstein limit~\cite{Boulware:1985wk,Wheeler:1985nh}. At particular values of the coupling, the two vacua can coincide. Such degeneracies also arise in the broader Lovelock family and, for suitable choices of couplings in odd dimensions, are associated with Chern--Simons formulations of gravity~\cite{Troncoso:1999pk,Crisostomo:2000bb}. This vacuum structure provides a means of exploring gravitational dynamics beyond the Einstein regime. This theory is also relevant to holography, where higher-curvature interactions modify the relation between bulk couplings and observables in the dual conformal field theory. In five bulk dimensions, the Gauss--Bonnet coupling controls the difference between the two central charges of the four-dimensional boundary CFT, providing a simple gravitational realization of $a\neq c$~\cite{Nojiri:1999mh,Blau:1999vz,Buchel:2009sk}. It also modifies the transport properties of the holographic plasma, notably the shear-viscosity-to-entropy-density ratio~\cite{Brigante:2007nu,Brigante:2008gz,Kats:2007mq}. Requiring causality and positivity of energy in the boundary theory imposes bounds on the allowed Gauss--Bonnet coupling~\cite{Brigante:2008gz,Hofman:2008ar,Buchel:2009tt,Camanho:2009vw}. EGBAdS gravity thus offers a tractable framework for relating consistency conditions in the boundary quantum field theory to constraints on higher-curvature interactions in the bulk.

\section{Taub--NUT--AdS with Gauss--Bonnet corrections}\label{sec:TN-EGB}

We consider higher-dimensional Taub--NUT geometries constructed as $U(1)$ fibrations over Einstein--K\"ahler base manifolds. Specifically, we study a family of inhomogeneous geometries constructed over complex line bundles~\cite{Page:1985bq,Corral:2025yvr}, with line element
\begin{equation}\label{ds-Taub-NUT}
    \dd s^2=-f(r)(\dd t+2n\mathcal{B})^2
    +\frac{\dd r^2}{f(r)}
    +(r^2+n^2)\dd\Sigma^2_{d-1}\,,
\end{equation}
where $t$ is the time coordinate, $r\in\mathbb{R}$ is the radial coordinate, and $n$ is the NUT parameter. The metric $\dd\Sigma^2_{d-1}$ describes a $(d-1)$-dimensional Einstein--K\"ahler manifold whose K\"ahler form is locally given by $\Omega=\dd\mathcal{B}$, with $\mathcal{B}$ a local potential one-form.

Taub--NUT geometries in EGB gravity were studied in Ref.~\cite{Dehghani:2005zm} and subsequently generalized to arbitrary Einstein--K\"ahler base manifolds and Lovelock theories at arbitrary order in the curvature in Ref.~\cite{Corral:2025yvr}. The metric function satisfying the field equations~\eqref{eom} is given by
\begin{widetext}
\begin{equation}\label{fsol}
    f_{\pm}(r)=\frac{\left[(d-1)^2(r^2+n^2)+2\Tilde{\alpha}\bar{R}\right](r^2+n^2)}
    {2\Tilde{\alpha}(d-1)\left[(d-2)^2-3n^2\right]}
   \left[
    1\pm\sqrt{
    1-\frac{4\Tilde{\alpha}r(d-1)^2
    \left[(d-2)r^2-3n^2\right][\mu(r)-m]}
    {\left[(d-1)^2(r^2+n^2)+2\Tilde{\alpha}\bar{R}\right]
    (r^2+n^2)^\frac{d-1}{2}}
    }\right]\,,
\end{equation}
\end{widetext}
where we have defined
\begin{equation}\label{musol}
    \mu(r)
    \equiv \frac{d(d-1)}{\ell^2}W^{(0)}(r)
    +\bar{R}W^{(1)}(r)
    +\alpha\bar{\mathcal{G}}W^{(2)}(r)\,,
\end{equation}
with
\begin{equation}
    \Tilde{\alpha}\equiv\alpha(d-1)(d-3)\label{alpha-tilde}\,,
\end{equation}
\begin{equation}
\begin{split}
    W^{(p)}(r)
    &\equiv\int^r
    \frac{(\rho^2+n^2)^{\frac{d+1}{2}-p}}{\rho^2}\dd\rho\\
    &=-\frac{(r^2+n^2)^{\frac{d+3-2p}{2}}}{n^2r}
    \,{}_2F_1\left(
    1,\frac{d+2-2p}{2};\frac{1}{2};-\frac{r^2}{n^2}
    \right)\,.
\end{split}
\end{equation}
Here, ${}_2F_1(a,b;c;z)$ is the Gauss hypergeometric function, $m$ is an integration constant, and $\bar{R}$ and $\bar{\mathcal{G}}$ denote the scalar curvature and Gauss--Bonnet density of the Einstein--K\"ahler base manifold, respectively. The parameter $\Tilde{\alpha}$ is the rescaled Gauss--Bonnet coupling.

At large $r$, the metric function~\eqref{fsol} behaves as
\begin{equation}\nonumber
\begin{split}
    f_\pm(r)
    &\simeq
    \frac{\left(\ell\pm\sqrt{\ell^2-4\alpha(d-2)(d-3)}\right)r^2}
    {2\alpha\ell(d-2)(d-3)}
    \\
    &\quad+
    \frac{(2d-1)\left[\ell\pm\sqrt{\ell^2-4\alpha(d-2)(d-3)}\right]n^2}
    {2\alpha\ell(d-2)^2(d-3)}\\
    &~~+\frac{\bar{R}}{(d-1)(d-2)} +\mathcal{O}(r^{-2})\,.
\end{split}
\end{equation}
Thus, the leading asymptotic term defines an effective AdS curvature radius,
\begin{equation}
    \left(\ell_{\rm eff}^\pm\right)^2=\frac{\ell^2}{2}\left(1\pm\sqrt{1-\frac{4\alpha(d-2)(d-3)}{\ell^2}}\right)\,.
\end{equation}
Each root corresponds to a distinct maximally symmetric vacuum of the theory, and the field equations~\eqref{eom} can be rewritten in the factorized form
\begin{equation}
\begin{split}
    \delta^{\mu\mu_1\cdots \mu_4}_{\nu\nu_1\cdots \nu_4}&\left(R^{\nu_1\nu_2}_{\mu_1\mu_2}+\frac{1}{(\ell_{\rm eff}^+)^2}\delta^{\nu_1\nu_2}_{\mu_1\mu_2}\right)\\
    &\times\left(R^{\nu_3\nu_4}_{\mu_3\mu_4}+\frac{1}{(\ell_{\rm eff}^-)^2}\delta^{\nu_3\nu_4}_{\mu_3\mu_4}\right)=0\,.
\end{split}
\end{equation}
Accordingly, the asymptotic line element is given by
\begin{equation}\label{ds-infty}
    \dd s^2\simeq
    \frac{(\ell_{\rm eff}^\pm)^2}{r^2}\dd r^2
    +r^2\left[
    -\frac{1}{(\ell_{\rm eff}^\pm)^2}(\dd t+2n\mathcal{B})^2
    +\dd\Sigma^2_{d-1}
    \right]\,.
\end{equation}
The NUT parameter remains encoded in the time fibration at infinity, so the asymptotic geometry does not, in general, coincide with that of global AdS$_{d+1}$. Nevertheless, the Riemann tensor approaches the constant-curvature form
\begin{equation}
    R^{\mu\nu}_{\lambda\rho}
    \simeq
    -\frac{2}{(\ell_{\rm eff}^\pm)^2}
    \delta^{[\mu}_{[\lambda}\delta^{\nu]}_{\rho]}\,.
\end{equation}

The two branches in Eq.~\eqref{fsol} have distinct limits as the Gauss--Bonnet coupling vanishes. The negative branch reduces to the Einstein--AdS solution~\cite{Awad:2000gg} as $\alpha\to0$,
\begin{widetext}
\begin{equation}
    f_-(r)\simeq
    \frac{r(r^2+n^2)^{\frac{d-1}{2}}}{d-1}
    \left[
    \frac{d(d-1)}{\ell^2}W^{(0)}(r)
    +\bar{R}W^{(1)}(r)-m
    \right]
    +\mathcal{O}(\alpha)\,, 
\end{equation}
\end{widetext}
whereas the positive branch behaves as
\begin{widetext}
\begin{equation}
\begin{split}
    f_+(r)
    &\simeq
    \frac{(r^2+n^2)^2}
    {\alpha(d-3)[(d-2)r^2-3n^3]}
    +\frac{2\bar{R}(r^2+n^2)}
    {(d-1)[(d-2)r^2-3n^2]}\\
    &~~-
    \frac{r}{(d-1)(r^2+n^2)^\frac{d-1}{2}}
    \left[
    \frac{d(d-1)}{\ell^2}W^{(0)}(r)
    +\bar{R}W^{(1)}(r)-m
    \right]
    +\mathcal{O}(\alpha)\,.
\end{split}
\end{equation}
\end{widetext}
Then, we see that the positive branch has no smooth Einstein limit, so we restrict our analysis to the negative branch and henceforth we consider $f(r)\equiv f_-(r)$. The existence of branches with no smooth Einstein limit is a broader feature of Lovelock theories; see Ref.~\cite{Garraffo:2008hu} for a review.

The spacetime admits the stationary Killing vector $\xi=\partial_t$. Isometries of the base manifold also extend to spacetime symmetries whenever they preserve the connection $\mathcal{B}$ up to a gauge transformation that can be compensated by a shift of the time coordinate.

\section{Thermodynamics}\label{sec:Thermo}

We seek a first law of full cohomogeneity in which the NUT parameter $n$ varies independently of the horizon radius, following the prescription of Hennigar, Kubiz\v{n}\'ak, and Mann~\cite{Hennigar:2019ive}. To this end, we consider the Lorentzian Taub--NUT metric given in Eqs.~\eqref{ds-Taub-NUT} and~\eqref{fsol}, without performing the usual Euclidean continuation, and retain the Misner string whenever it is present. In the four-dimensional Lorentzian Taub--NUT geometry, Refs.~\cite{Clement:2015aka,Clement:2015cxa} showed that retaining the Misner string does not obstruct geodesic completeness and that freely falling observers do not encounter closed causal geodesics. These results, however, do not imply the absence of arbitrary closed timelike curves or establish a globally well-behaved causal structure. Moreover, their direct extension to the higher-dimensional EGB configurations considered here has not been demonstrated. In the present work, we therefore retain the Misner string as part of the Lorentzian thermodynamic prescription of Ref.~\cite{Hennigar:2019ive}, without making stronger claims concerning the global causal structure of the full family of solutions. Consequently, the NUT parameter can vary freely in the space of solutions.

The most straightforward thermodynamic quantities to compute are those determined by the horizon geometry. The outer horizon is located at $r=r_+$, where $r_+$ is the largest real root of $f(r_+)=0$. This condition allows us to express the integration constant $m$ in terms of the horizon radius as
\begin{equation}\label{mrp}
    m=\mu(r_+)\,,
\end{equation}
where $\mu(r)$ is defined in Eq.~\eqref{musol}. The Hawking temperature is obtained in the usual way from the surface gravity $\varkappa$, defined by
\begin{equation}
    \varkappa^2=\eval{-\frac{1}{2}(\nabla_\mu\xi_\nu)(\nabla^\mu\xi^\nu)}_{r=r_+}\,,
\end{equation}
where $\xi=-\partial_t$ is the stationary Killing vector generating the horizon. A direct computation yields
\begin{equation}
\begin{split}
    T=\frac{\varkappa}{2\pi}
    =\frac{(d-1)r_+}{4\pi\left(r_+^2+n^2\right)^\frac{d-3}{2}}
    \left|\frac{\mu'(r_+)}{2\Tilde{\alpha}\bar{R}+(d-1)^2\left(r_+^2+n^2\right)}\right|\,.
\end{split}
\end{equation}

To compute the energy of these spacetimes in arbitrary dimensions, we employ the off-shell ADT formalism introduced in Ref.~\cite{Kim:2013zha}, which provides a prescription for constructing conserved charges in generic theories of pure gravity from an off-shell ADT current. Building on Refs.~\cite{Barnich:2001jy,Barnich:2003xg,Barnich:2004uw}, it incorporates an integration along a path in the space of solutions, described by one or more continuous solution parameters. This construction offers a systematic method for evaluating conserved quantities without introducing explicit boundary counterterms. The formalism has been applied to quasi-local conserved charges in a broad class of gravitational theories and extended to include various matter sectors~\cite{Hyun:2014sha,Peng:2014gha,Setare:2016dex,Peng:2016eyx,Peng:2016qnz}.

Within this prescription, the conserved charge associated with a Killing vector $\xi=\xi^\mu\partial_\mu$ is
\begin{equation}\label{Q-xi}
    \Q[\xi]
    =
    \int_\Sigma
    \dd\Sigma_{\mu\nu}
    \left(
        \Delta K^{\mu\nu}(\xi)
        -2\xi^{[\mu}
        \int_0^1\dd s\,\Theta^{\nu]}
    \right)\,,
\end{equation}
where
\begin{equation}
    \Delta K^{\mu\nu}(\xi)
    \equiv
    K^{\mu\nu}_{s=1}(\xi)
    -
    K^{\mu\nu}_{s=0}(\xi)
\end{equation}
is the difference between the Noether potentials evaluated on the solution of interest and the reference background, respectively. Here, $s$ parametrizes the interpolating path, and $\dd\Sigma_{\mu\nu}$ is the integration element on a codimension-two surface $\Sigma$. For the EGBAdS action in Eq.~\eqref{I-EGB}, the surface term and Noether potential are given by
\begin{equation}\label{Theta-K}
\begin{aligned}
    \Theta^\mu
    &=2P^{\mu\alpha\beta\gamma}\nabla_\gamma\delta g_{\alpha\beta}
    -\delta g_{\alpha\beta}\nabla_\gamma P^{\mu\alpha\beta\gamma}\,,
    \\
    K^{\mu\nu}
    &=2P^{\mu\nu\rho\sigma}\nabla_\rho\xi_\sigma
    -4\xi_\sigma\nabla_\rho P^{\mu\nu\rho\sigma}\,,
\end{aligned}
\end{equation}
respectively, where $P^{\mu\nu\rho\sigma}$ is defined as the derivative of the Lagrangian with respect to the Riemann tensor, which for the action in Eq.~\eqref{I-EGB}, it takes the form
\begin{equation}\label{P-tensor}
    P^{\mu\nu}_{\alpha\beta}
    =\frac{\kappa}{2}\left(
    \delta^{\mu\nu}_{\alpha\beta}
    +\alpha\delta^{\mu\nu\lambda\rho}_{\alpha\beta\tau\sigma}
    R^{\tau\sigma}_{\lambda\rho}
    \right)\,.
\end{equation}
The formula in Eq.~\eqref{Q-xi} defines a quasi-local conserved charge. For certain classes of spacetimes, this charge is independent of the radial coordinate and may therefore be evaluated on any suitable codimension-two surface~\cite{Kim:2013zha}. More generally, the asymptotic charge is obtained by taking the limit
\begin{equation}\label{Q-inf}
    \Q^\infty[\xi]
    =
    \lim_{r\to\infty}\Q[\xi]\,.
\end{equation}
Notice that in the present theory, the tensor $P^{\mu\nu\rho\sigma}$ is identically divergence-free, so the terms involving its covariant derivatives vanish in both $\Theta^\mu$ and $K^{\mu\nu}$.

It is worth mentioning that a limitation arises when applying this linearization-based method to Lovelock theories with a unique degenerate AdS vacuum, obtained through a particular choice of couplings~\cite{Crisostomo:2000bb,Kastor:2006vw}. Around such a vacuum, the linearized field equations vanish identically, so the standard linearized construction does not provide a nontrivial conserved charge. Consequently, a propagating graviton cannot be identified within the usual linearized approximation around that background~\cite{Camanho:2009hu,Camanho:2010ru}. As a matter of fact, the holographic implications of the degeneracy are not well understood and further insights of its role is needed. See Refs.~\cite{Arenas-Henriquez:2017xnr,Arenas-Henriquez:2019rph} for a discussion of the definition of energy in these theories.

We now apply this prescription to compute the energy of the Taub--NUT AdS configurations. As can be seen from Eq.~\eqref{ds-infty}, the presence of the NUT parameter prevents the metric from being asymptotically AdS$_{d+1}$ in the strict sense. Therefore, the asymptotic geometry is defined by setting $m=0$ while keeping $n$ fixed. A one-parameter path in the space of solutions is then defined by the replacement $m\mapsto s\,m$, with $s\in[0,1]$, interpolating continuously between the asymptotic geometry at $s=0$ and the configuration of interest at $s=1$. For the metric in Eqs.~\eqref{ds-Taub-NUT} and~\eqref{fsol}, the components of the surface term and Noether potential relevant to the energy are
\begin{widetext}
\begin{equation}
\begin{split}
    \Theta^r&=-\kappa\dv{m}\left(f'+\frac{(d-3)r}{r^2+n^2}f+\frac{2\alpha(d-1)}{(r^2+n^2)^3}\left.\Big\{(r^2+n^2)[3n^2-(d-2)r^2]ff'\right.\right.\\
    &\left.\left.\quad-\frac{r}{2}[(11-d)n^2+(d-2)(d-3)r^2]f^2
    +\frac{\bar{R}}{d-1}(r^2+n^2)[(r^2+n^2)f'+(d-3)rf]\right\}\right)\,,
\end{split}
\end{equation}
and
\begin{equation}
\begin{split}
    K^{tr}(-\partial_t)&=\kappa f'
    +\frac{2\kappa\alpha}{(r^2+n^2)^3}
    \left\{\bar{R}(r^2+n^2)^2f'
    +(d-1)(r^2+n^2)[5n^2-(d-2)r^2]ff'-4(d-1)n^2rf^2\right\}\,,
\end{split}
\end{equation}
\end{widetext}
respectively, where $f=f(r)$, and prime denotes differentiation with respect to the radial coordinate $r$. We notice that the surface term can be written as a total derivative with respect to $m$, which makes the integration along the solution path straightforward. Evaluating Eq.~\eqref{Q-inf} then gives
\begin{equation}\label{Mass}
    \Q^\infty[-\partial_t]\equiv M
    =\frac{\Sigma_{(d-1)}}{16\pi G}m\,,
\end{equation}
confirming that the integration constant $m$ plays the role of a mass parameter.

The entropy can also be obtained within this framework through its interpretation as the Noether charge associated with the appropriately normalized horizon-generating Killing vector~\cite{Wald:1993nt}. For the horizon generated by $\xi=-\partial_t$, the term proportional to $\xi$ in Eq.~\eqref{Q-xi} vanishes on the bifurcation surface, leaving only the contribution of the Noether potential. Evaluating the expression~\eqref{Q-xi} at $r=r_+$ yields
\begin{equation}\label{S-nut}
    S=\frac{\Sigma_{(d-1)}(r_+^2+n^2)^\frac{d-1}{2}}{4G}
    \left(1+\frac{2\alpha\bar{R}}{r_+^2+n^2}\right)\,,
\end{equation}
where $\Sigma_{(d-1)}$ denotes the volume of the transverse Einstein--K\"ahler base manifold.

This result agrees with the entropy obtained from the Iyer--Wald formula~\cite{Wald:1993nt,Iyer:1994ys,Iyer:1995kg}, given by
\begin{equation}\label{S-Wald}
    S_{\rm IW}
    =-2\pi\int_{\mathcal H}\sqrt{|h|}\dd^{d-1}x\,
    P^{\mu\nu\alpha\beta}
    \hat{\epsilon}_{\mu\nu}\hat{\epsilon}_{\alpha\beta}\,,
\end{equation}
where $\hat{\epsilon}_{\mu\nu}$ is the binormal to the bifurcation surface $\mathcal H$, normalized such that $\hat{\epsilon}_{\mu\nu}\hat{\epsilon}^{\mu\nu}=-2$, and $h$ is the determinant of the induced metric. The tensor $P^{\mu\nu\alpha\beta}$, defined in Eq.~\eqref{P-tensor}, encodes the corrections to the Bekenstein--Hawking entropy arising from higher-curvature interactions or nonminimal couplings.

Since $\mathcal{H}$ is a surface of constant $t$ and $r$, only the component $P^{trtr}$ contributes to the contraction with the binormals in Eq.~\eqref{S-Wald}. This component is given by
\begin{equation}
    P^{trtr}=\frac{\kappa}{2}
    +\alpha\kappa\left\{
    \frac{\bar{R}}{r^2+n^2}
    +\frac{(d-1)f}{(r^2+n^2)^2}
    \left[3n^2-(d-2)r^2\right]
    \right\}\,.
\end{equation}
Then, evaluating Eq.~\eqref{S-Wald} at $r=r_+$ yields the same result as Eq.~\eqref{S-nut}, confirming the agreement between the two approaches. This is expected, since the quasilocal off-shell ADT construction is equivalent to the covariant phase space formalism underlying Wald's entropy prescription~\cite{Kim:2013cor,Kim:2013zha}. We notice that the Gauss--Bonnet interaction corrects the Bekenstein--Hawking area law, as expected in Lovelock gravity~\cite{Jacobson:1993xs,Cai:2001dz}

Since $n$ is no longer constrained to be a function of the horizon radius, its independent variation must be accommodated in the first law. To obtain a first law of full cohomogeneity, we introduce an additional thermodynamic variable associated with the independent variation of the NUT parameter, following Ref.~\cite{Hennigar:2019ive}. This construction is motivated by the thermodynamics of accelerating black holes, where an additional term accounts for the tension of the string responsible for their acceleration~\cite{Appels:2017xoe,Anabalon:2018ydc,Anabalon:2018qfv}. Thus, the first law takes the form
\begin{equation}\label{first-law}
    \delta M=T\delta S+\psi\delta N\,,
\end{equation}
where $N$ denotes the extensive NUT-related thermodynamic variable, while $\psi$ is its conjugate potential. One could also allow the cosmological constant to vary and introduce a conjugate thermodynamic volume~\cite{Hennigar:2019ive}. Here, however, we keep the cosmological constant fixed, as it does not arise as an integration constant in the present construction. Allowing it to vary would also require treating the Gauss--Bonnet coupling $\alpha$ as a thermodynamic variable and introducing its conjugate quantity, whose physical interpretation is unclear.

Assuming that $\psi$ depends only on $n$, the integrability conditions associated with Eq.~\eqref{first-law} imply $\psi\propto 1/n$. Direct integration then yields
\begin{widetext}
\begin{equation}\label{psi-N}
    \psi=\frac{c}{n}\quad \text{and}\quad
    N=\frac{\Sigma_{(d-1)}n^3}{16\pi c G}\left[\frac{d(d-1)}{\ell^2}W^{(1)}(r_+)
    +\bar{R}W^{(2)}(r_+)
    +\alpha\bar{\mathcal{G}}W^{(3)}(r_+)\right]\,,
\end{equation}
\end{widetext}
where $c$ is a nonzero integration constant that fixes the normalization of the thermodynamic pair $(\psi,N)$. Notice that the thermodynamic charge $N$ involves the same combination of curvature invariants as $\mu(r)$ in Eq.~\eqref{musol}, but with each polynomial $W^{(p)}(r)$ replaced by $W^{(p+1)}(r)$ and evaluated at the horizon.

We emphasize that \(N\) is defined here thermodynamically through the integrability of the full-cohomogeneity first law. More precisely, under the assumption that the conjugate potential depends only on the NUT parameter, \(\psi=\psi(n)\), the integrability conditions determine the functional dependence of \(\psi\) and the corresponding extensive variable \(N\). Thus, Eq.~\eqref{psi-N} establishes \(N\) as a thermodynamically consistent NUT-related quantity, but does not by itself provide an independent geometrical definition as a Noether--Wald or Misner-string charge. Importantly, this thermodynamic construction does not require the presence of a Misner string and applies uniformly to the family of Einstein--K\"ahler base manifolds considered here. Since the existence of Misner strings depends on the topology of the base, a geometrical construction based on tubular surfaces surrounding the strings, while potentially illuminating for the corresponding subclass of solutions, would not provide a universal definition of \(N\). It would nevertheless be interesting to determine whether, whenever Misner strings are present, such a construction reproduces \(N\), or the corresponding combination \(\psi N\), obtained thermodynamically, as occurs in the four-dimensional Einstein case~\cite{Bordo:2019tyh}.

The Einstein--AdS limit of these quantities follows directly by taking $\alpha\to0$, yielding a higher-dimensional generalization of the results of Ref.~\cite{Hennigar:2019ive} with the cosmological constant held fixed rather than treated as a thermodynamic pressure.

\subsection{Static limit}\label{sec:Static-limit}
The static limit of the geometries considered above is obtained by setting the NUT parameter to zero, $n\to0$. The line element then becomes
\begin{equation}
    \dd s^2=-f(r)\dd t^2+\frac{\dd r^2}{f(r)}
    +r^2\dd\Sigma^2_{d-1}\,,
\end{equation}
where the metric function solving the field equations~\eqref{eom} is given by
\begin{widetext}
\begin{equation}
    f_{\pm}(r)
    =
    \frac{(d-1)^2r^2+2\tilde{\alpha}\bar{R}}
    {2\tilde{\alpha}(d-1)(d-2)}
    \left[
    1\pm\sqrt{
    1-\frac{4\tilde{\alpha}(d-1)(d-2)[\mu(r)-m]}
    {[(d-1)r^2+2\tilde{\alpha}(d-3)\bar{R}]^2r^{d-4}}
    }
    \right]\,,
\end{equation}
\end{widetext}
with
\begin{equation}\label{mu-static}
    \mu(r)
    =
    \frac{(d-1)r^d}{\ell^2}
    +\frac{\bar{R}r^{d-2}}{d-2}
    +\alpha\frac{\bar{\mathcal{G}}r^{d-4}}{d-4}\,.
\end{equation}
This general family of solutions was studied in Ref.~\cite{Ray:2015ava} and includes, for suitable choices of the transverse geometry, several previously known solutions~\cite{Khodam-Mohammadi:2008hww,Cai:2001dz}. As before, only the negative branch approaches the corresponding Einstein-gravity solution continuously as $\alpha\to0$, and we therefore restrict our analysis to this branch.

The horizon condition determines the mass parameter through $m=\mu(r_+)$, as in Eq.~\eqref{mrp}, with $\mu(r)$ now given by Eq.~\eqref{mu-static}. Taking the corresponding limits of the thermodynamic quantities derived above yields the temperature, entropy, and energy,
\begin{subequations}
\begin{align}
    T&=\frac{(d-1)r_+}{4\pi r_+^{d-3}}
    \left|
    \frac{\mu'(r_+)}
    {2\tilde{\alpha}\bar{R}+(d-1)^2r_+^2}
    \right|\,,
    \\
    S&=\frac{\Sigma_{(d-1)}r_+^{d-1}}{4G}
    \left(1+\frac{2\alpha\bar{R}}{r_+^2}\right)\,,
    \\
    M&=\frac{\Sigma_{(d-1)}}{16\pi G}m\,,
\end{align}
\end{subequations}
respectively. With the couplings and transverse geometry held fixed, these quantities satisfy the standard first law,
\begin{equation}
    \delta M=T\delta S\,.
\end{equation}
identically. For $d=5$, the mass agrees with the result obtained in Ref.~\cite{Anastasiou:2025usa} using holographic renormalization in EGBAdS gravity. The limit $\alpha\to0$ can be taken directly in these expressions, smoothly recovering the corresponding thermodynamic quantities in Einstein--AdS gravity for topological black holes in higher dimensions.

\section{Discussion}\label{sec:Conclusions}
In this work, we have investigated the thermodynamics of Taub--NUT spacetimes in EGBAdS gravity, following the prescription of Hennigar, Kubiz\v{n}\'ak, and Mann in Ref.~\cite{Hennigar:2019ive}. The conventional treatment of NUT-charged spacetimes relies on a Euclidean continuation and imposes a relation between the NUT parameter and the horizon radius to eliminate the Misner string. Here, we instead work in Lorentzian signature and retain the Misner string, since the spacetime remains geodesically complete even in its presence~\cite{Clement:2015aka,Clement:2015cxa} . This approach allows the NUT parameter and the horizon radius to vary independently and motivates the construction of a homogeneous first law. To this end, the standard first law must be supplemented by a thermodynamic charge associated with the NUT parameter and its conjugate potential, as shown in Eq.~\eqref{first-law}. Although the AdS radius can also be promoted to a thermodynamic variable, as in Ref.~\cite{Hennigar:2019ive}, we keep it fixed. An extended thermodynamic treatment would also require allowing the Gauss–Bonnet coupling $\alpha$ to vary and introducing its conjugate quantity, whose physical interpretation is less clear. We computed the energy using the off-shell ADT formalism and found it to be proportional to the integration constant $m$, as expected. The entropy was also obtained within this prescription as a horizon charge and agrees with the result derived from the Iyer--Wald formula~\cite{Iyer:1994ys,Iyer:1995kg}. The latter  receives a Gauss–Bonnet correction to the Bekenstein–Hawking area law of general relativity. With these quantities, we constructed the first law and explicitly determined the additional NUT-related charge and its conjugate potential required for it to hold. We stress that \(N\) is defined here as a thermodynamic variable and should not, in general, be identified with a Misner-string charge, since such a geometrical construction is not available for all base topologies considered in the present family. For the subclass admitting Misner strings, however, the geometrical construction of Ref.~\cite{Bordo:2019tyh} suggests an interesting complementary characterization.

We also examined the static limit, in which our thermodynamic quantities agree with those obtained through holographic renormalization for the six-dimensional topological Boulware–Deser black hole~\cite{Anastasiou:2025usa}. We expect the higher-dimensional extensions of our results to agree with the corresponding holographic renormalization calculations, for which the number and complexity of the required counterterms generally increase with the spacetime dimension.

Several directions remain open for further investigation. A particularly interesting extension is to the full Lovelock theory, for which solutions with a NUT parameter are known in closed form~\cite{Corral:2025yvr}. The derivative of the Lovelock Lagrangian with respect to the curvature can be obtained directly from the action in Eq.~\eqref{I-Lovelock}. The mass could then be computed by evaluating the surface term and Noether potential entering Eq.~\eqref{Q-inf}, and we expect it to retain a relation of the form given in Eq.~\eqref{Mass}. Likewise, the entropy could be derived by following the procedure used to obtain Eq.~\eqref{S-nut}. This would extend known results for black hole thermodynamics in Lovelock gravity~\cite{Jacobson:1993xs,Myers:1988ze,Cai:2003kt,Crisostomo:2000bb,Kastor:2010gq,Cai:2001dz,Frassino:2014pha,Wu:2025xxo} to the stationary case. We address this extension in a forthcoming work.

Another direction is to investigate how electromagnetic fields and angular momentum modify the first law. The latter can be introduced through an improper diffeomorphism that changes the distribution of the Misner string between the north and south poles; see, for example, Ref.~\cite{Durka:2019ajz}. More broadly, it would be worthwhile to study the thermodynamics of Lorentzian NUT-charged solutions in theories beyond general relativity, including those with matter fields~\cite{Corral:2021xsu,Corral:2025npd,Corral:2024lva}, and to compare the resulting thermodynamic descriptions with conventional Euclidean treatments that eliminate the Misner string. We leave these questions for future work.

\begin{acknowledgments}
   The author is deeply grateful to Crist\'obal Corral for his insightful comments and constructive criticism on an earlier version of this manuscript, as well as for numerous stimulating discussions on these topics over the years. He is also grateful to Eloy Ay\'on-Beato for encouraging him to undertake this calculation. He further thanks Gustavo Aldaz, Luis Avil\'es, Benjam\'{i}n Hern\'andez, and Sim\'on del Pino for valuable discussions.
\end{acknowledgments}

\bibliography{biblio}

\end{document}